\documentclass[11pt,twoside]{article}
\usepackage{asp2014}

\aspSuppressVolSlug
\resetcounters

\begin{document}

\title{Testing common envelopes on double white dwarf binaries}
\author{Jose L.\ A.\ Nandez$^1$, Natalia Ivanova$^2$, and James C.\ Lombardi, Jr.$^3$
\affil{$^1$Department of Physics, University of Alberta, Edmonton, AB, T6G 2E7, Canada; \email{avendaon@ualberta.ca}}
\affil{$^2$Department of Physics, University of Alberta, Edmonton, AB, T6G 2E7, Canada; \email{nata.ivanova@ualberta.ca}}
\affil{$^3$Department of Physics, Allegheny College, Meadville, PA 16335, USA; \email{jalombar@allegheny.edu}}}

\paperauthor{Jose L. A. Nandez}{avendaon@ualberta.ca}{0000-0001-6282-0539}{University of Alberta}{Department of Physics}{Edmonton}{Alberta}{T6G 2E7}{Canada}
\paperauthor{Natalia Ivanova}{nata.ivanova@ualberta.ca}{}{University of Alberta}{Department of Physics}{Edmonton}{AB}{T6G 2E7}{Canada}
\paperauthor{James C. Lombardi Jr}{jalombar@allegheny.edu}{}{Allegheny College}{Department of Physics}{Meadville}{PA}{16335}{USA}

\begin{abstract}
The formation of a double white dwarf binary likely involves a common envelope (CE) event 
between a red giant  and a white dwarf (WD) during the most recent episode of Roche lobe overflow mass transfer.
We study the role of recombination energy with hydrodynamic simulations of such stellar interactions. 
We find that the recombination energy helps to expel the common envelope entirely, 
while if recombination energy is not taken into account, a significant fraction of the common envelope remains bound.
We apply our numerical methods to constrain the progenitor system for WD 1101+364 -- a double WD binary that has well-measured 
mass ratio of $q=0.87\pm0.03$  and an orbital period of 0.145 days. Our best-fit progenitor for the pre-common envelope donor is a 1.5 $M_\odot$ red giant. 
\end{abstract}

\section{Introduction}
\label{sec:intro}
The formation channel of a double white dwarf (DWD) binary is still not fully certain, although 
it is believed that the formation of a compact binary system composed of two white dwarfs 
(WDs) includes a common envelope (CE) event, 
at least during the last episode of mass exchange between the first-formed WD and a low-mass giant. 
Low-mass giants have a well-defined relation between their core masses and radii. 
This makes the state of the progenitor binary system at the onset of the CE event leading to a DWD 
the best theoretically understood from among the progenitors of all known types of post-CE systems.
However, previous attempts to model a CE event between a low-mass giant and a WD did not succeed to eject the entire CE during
three-dimensional hydrodynamical simulations -- a significant fraction of the expanded envelope remained bound to the formed binary, 
with almost no energy transfer taking place from the binary orbit to that circumbinary envelope  
\citep{2012ApJ...744...52P,2012ApJ...746...74R}. 
It has been proposed a while ago that the recombination energy of H and He should 
play a role in unbinding the donor's envelope 
\citep{1967AJ.....72Q.813L,1967Natur.215..838R,1968AcA....18..255P,1994MNRAS.270..121H,2002MNRAS.336..449H};
however, to date, this energy has not yet been taken into account in the modeling of a CE event.

\section{Numerical method}

We use a 3D Smoothed Particle Hydrodynamics (SPH) code, {\tt StarSmasher} \citep{2010MNRAS.402..105G,2011ApJ...737...49L};
for details on modeling of CE events see \cite{0004-637X-786-1-39}.
This code by default uses a standard analytical SPH equations of state (EOS) that takes 
into account radiation pressure and ideal gas pressure, but not the changes in gas' ionization states.
We added a possibility to run the code using a tabulated EOS adapted from {\tt MESA} \citep{2011ApJS..192....3P}, where recombination energy is taken into account. 
To model donor stars, we first evolve one-dimensional low-mass stars using {\tt EV/stars} \citep[recent updates described in][]{2008A&A...488.1007G} and then
relax them in {\tt StarSmasher}.
The comparison of the initial relaxed stars with two different EOSs have shown that SPH profiles in both cases 
match very well the pressure, density, and gravitational potential with the stellar profiles. 
However, the SPH specific internal energy profile is matched only in case when the tabulated EOS is used (see Figure 1). 

\articlefigure[scale=0.45]{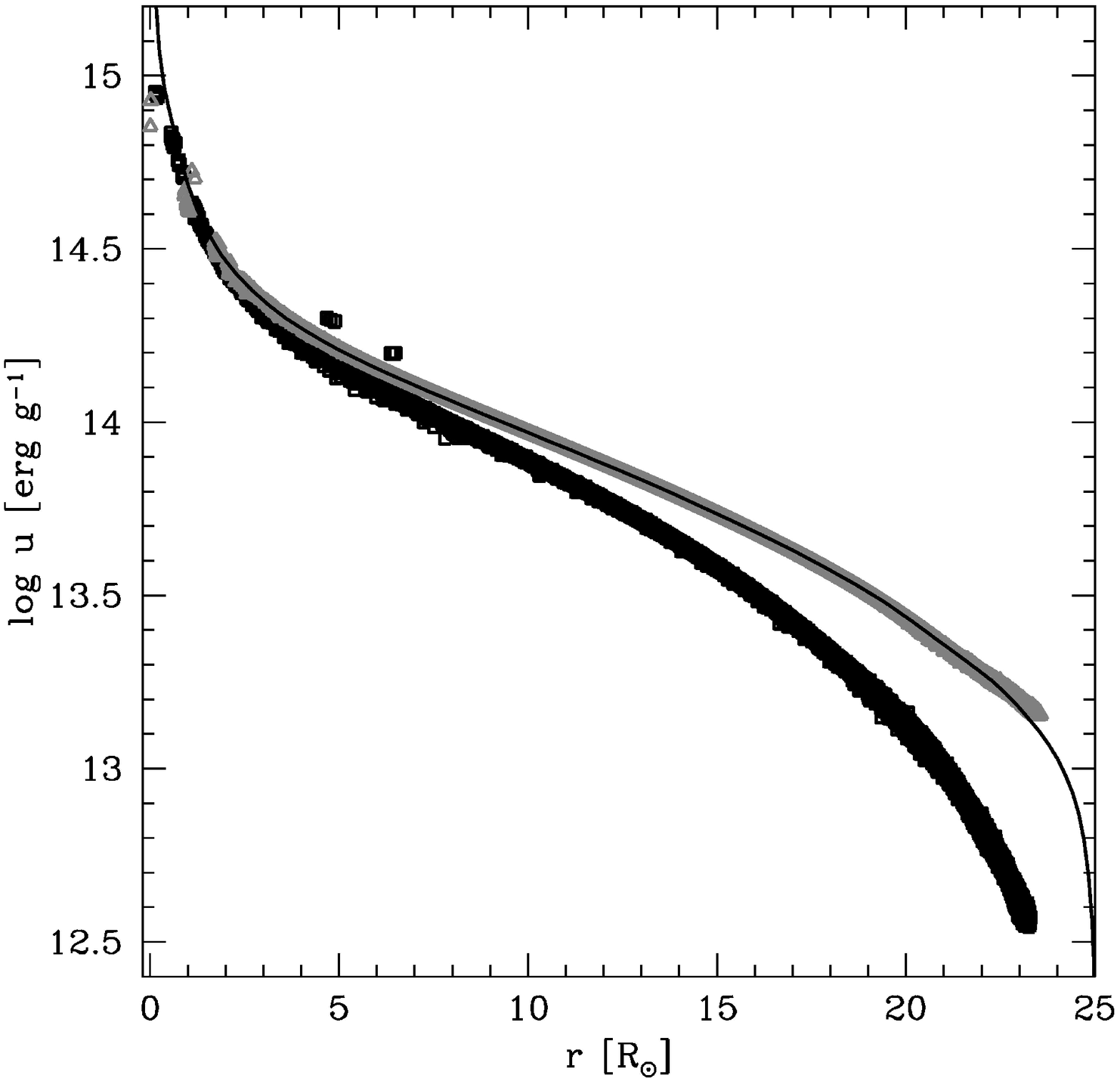}{fig:num2}{The specific internal energy $u$ as a function of the radius coordinate, for the 1.5$M_\odot$ giant with 0.31 $M_\odot$ core. The solid line corresponds to one-dimensional stellar profile, the gray triangles correspond to the relaxed star with the tabulated EOS, and the black boxes are for the standard EOS.}

\section{Initial conditions}

One of the best measured DWD systems is WD 1101+364.  This system has an orbital period of 0.145~d and the mass ratio between the younger and older WDs $q=m_1/m_2=0.87\pm0.03$, where $m_1=0.31M_\odot$ and $m_2=0.36M_\odot$  \citep{1995MNRAS.275L...1M}.
Accordingly, we adopted for our CE simulations that the WD companion has a mass of 0.36$M_\odot$. 
For a red giant companion, we considered low-mass stars with masses  1, 1.2, 1.4, 1.5, 1.7 and 1.8 $M_\odot$.
We have evolved those stars until the moment when the core has a He mass of $0.31-0.32M_\odot$.
Figure 2 shows how the stellar radius changes when the He core mass grows.
We adopted that the radius that a star has when its core mass is $0.31-0.32M_\odot$ is equal to the Roche lobe radius defined as in \cite{1983ApJ...268..368E}.
This provided us with the initial orbital separation of the progenitor binary at the start of our CE simulations.
Because the radii of the giants are a strong function of the core mass and not the total mass, the initial orbital separations 
do not vary strongly among the cases with giants of different masses.

\articlefigure[scale=0.45]{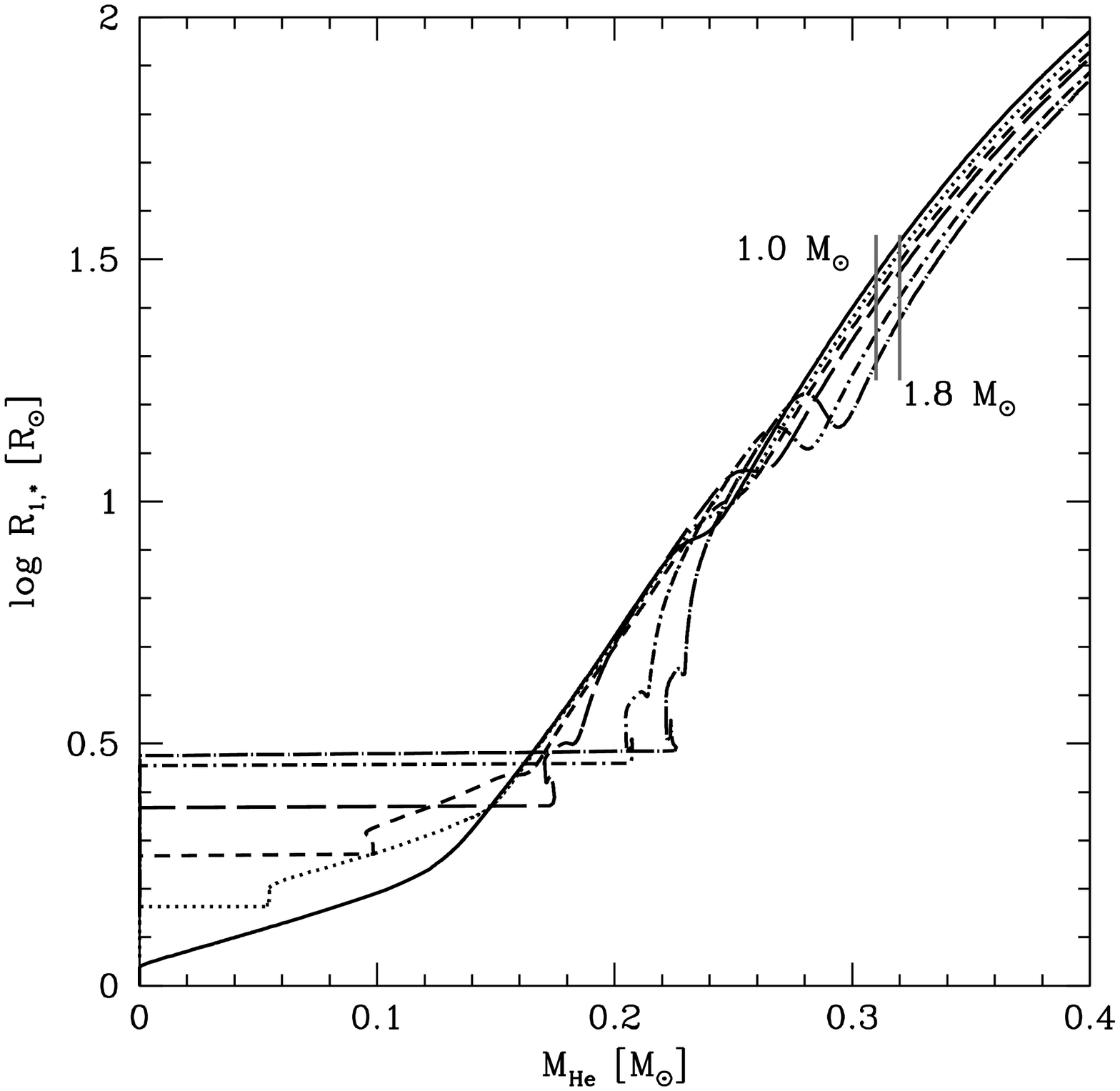}{fig:num1}{Stellar radii as a function of the red giant's He core mass. 
The stellar tracks correspond to 1.0, 1.2, 1.4, 1.5, 1.7 and 1.8 $M_\odot$. 
The gray vertical lines indicate the region of possible initial radii of the red giants, 
to be consistent with the observation of WD 1101+364.}

\section{Results}

We compare the two EOSs for the example of an initial binary consisting of 1.5$M_\odot$ giant star (modeled with 99955 particles) and a  
0.36$M_\odot$ WD (modeled as a point-mass). For each particle, we find its total energy as the sum of its kinetic, potential and internal energies (internal energy here
includes recombination energy, if still stored). We then define:

\begin{itemize}\itemsep0em
 \item {\it Ejecta} -- the unbound material where each particle has its total energy positive;
 \item {\it Binary} -- all the particles that have their total energy negative, and are located inside of one of the two Roche lobes;
 \item {\it Circumbinary} -- all particles which are bound to a formed binary (the total energy of each particles is negative), 
and are located outside of the either of the Roche lobes. 
\end{itemize}
The simulations were stopped  after 6000 orbital periods counted from the compact binary formation. 
See the Table~1 for the masses and total energies at the end of the simulations. 
In our simulations, just as in the previous studies  \citep{2012ApJ...744...52P,2012ApJ...746...74R}, the run with the standard EOS resulted in more than a half 
of red giant envelope to remain bound to the system, even though the value of its total energy is very small compared to the orbital energy of the binary. 
At the same time, the run that used the tabulated EOS resulted in the whole envelope being expelled. 

The energy formalism that described a CE event simplistically \citep[for an overview, see][]{2013A&ARv..21...59I} 
predicts that the more massive the donor is, the tighter the orbit gets. 
Indeed, we find the same trend with our simulations -- see Figure  3, where 
we show how the final orbital period changes with the initial giant's mass, 
only for the runs that were performed using the tabulated EOS.

Our best match with the observed orbital period $P=0.145$d \citep{1995MNRAS.275L...1M} is obtained for a donor of 1.5 M$_\odot$.

\articlefigure[scale=0.45]{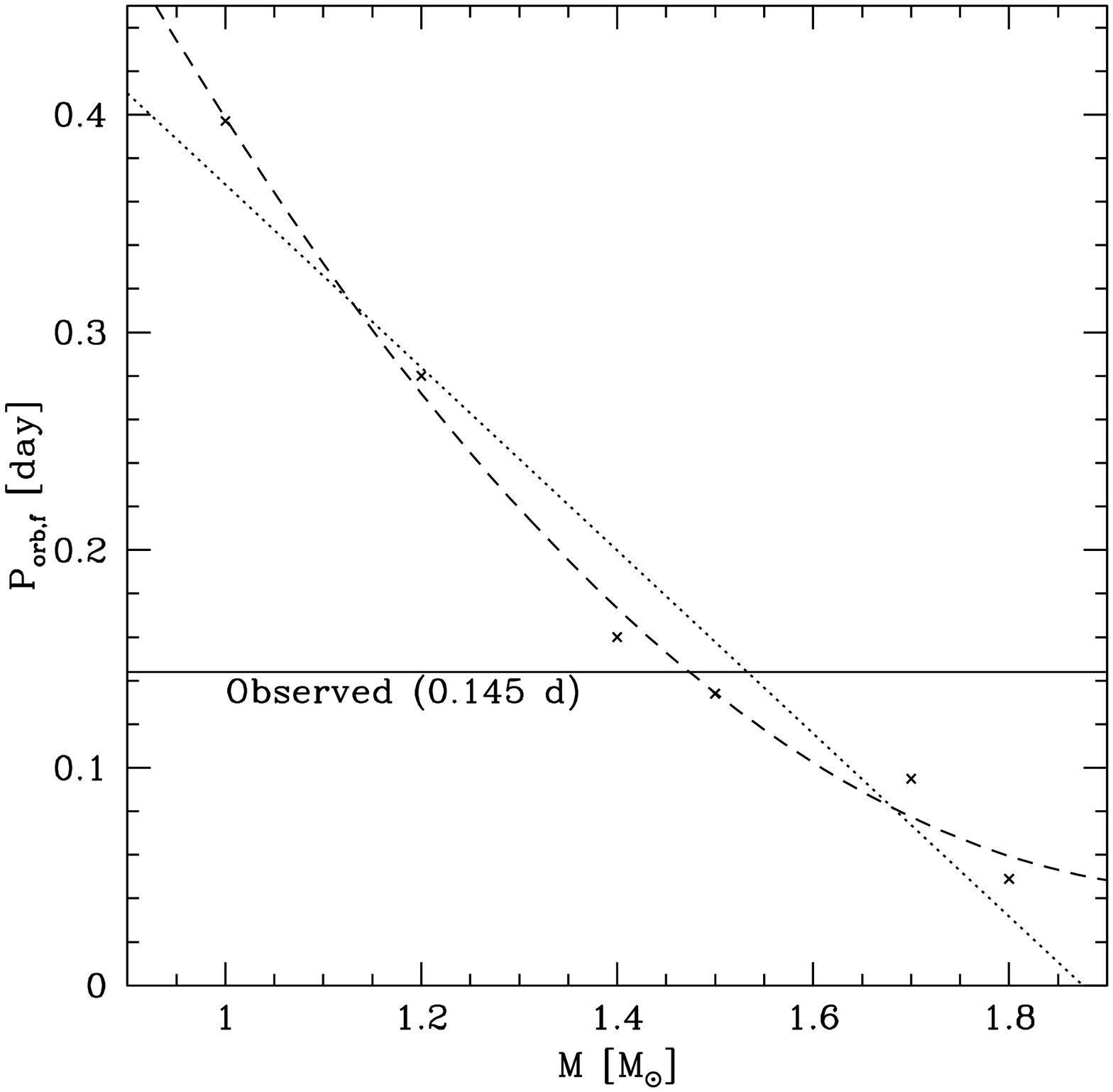}{fig:num3}{
Final orbital periods, as a function of the pre-CE giant mass.  The dotted and dashed curves represent, respectively, simple linear and quadratic fits to our simulation result.  The solid line represents the observed period of WD 1101+364.
}

\begin{table}
 \label{tab:num1}
 \caption{CE event involving 1.5$M_\odot$ giant and 0.36$M_\odot$ WD.}
 \begin{center}
 \begin{tabular}{cccccccccc}
  \hline
  EOS & $m_{\rm unb}$ & $m_{\rm cir}$ & $m_{\rm bin}$ & $E_{\rm tot,unb}^{\infty}$ & $E_{\rm tot,cir}^{\infty}$ & $E_{\rm tot,bin}^{\infty}$ \\
  \hline
  standard EOS & 0.521 & 0.658 & 0.679 & 2.150 & -0.402 & -24.011 \\
  tabulated EOS & 1.178 & 0.000 & 0.680 & 6.910 &  0.000 & -25.508 \\
  \hline
 \end{tabular}
 \end{center}
$m_{\rm unb}$ is the mass of the ejecta, 
$m_{\rm cir}$ is the mass of circumbinary material, 
and $m_{\rm bin}$ is the mass of the formed binary, all in $M_\odot$. 
$E_{\rm tot,unb}^{\infty}$, $E_{\rm tot,cir}^{\infty}$, $E_{\rm tot,bin}^{\infty}$ are the total energy of the ejecta, circumbinary material, and the binary, respectively, all in units of $10^{46}$ erg. 
\end{table}

\section{Conclusion}

With our simulations, we have confirmed that the recombination energy plays an important role in the ejection of a CE, where it helps to 
unbind the entire envelope.

The observations for WD 1101+364 are well reproduced with a progenitor binary composed 
of a 1.5$M_\odot$ red giant and a $0.36M_\odot$ WD companion.
In this case, the modeled post-CE binary has a final orbital period of $P_{\rm orb}=0.135$ d, the newly formed WD has a mass of $0.319M_\odot$ 
and an older WD has a mass $0.361M_\odot$; binary's mass ratio is $q=0.884$, in agreement with the observational error. 

\acknowledgements J.L.A.N.\ would like to thank the Mary Louise Imrie Graduate Student Award from the University of Alberta and CONACyT for their support in this project. N.I.\ acknowledges support by NSERC Discovery and the Canada Research Chairs program. J.C.L.\ is supported by the National Science Foundation (NSF)
Grant No. AST-1313091.

\bibliography{JNandez}

\end{document}